# Assimilation of SAR-derived flood extent observations for improving fluvial flood forecast – A proof-of-concept

**Thanh Huy Nguyen**[1], **Sophie Ricci**[1], **Andrea Piacentini**[1], **Christophe Fatras**[2], **Peter Kettig**[3], **Gwendoline Blanchet**[3], **Santiago Peña Luque**[3], **and Simon Baillarin**[3]

[1] CECI, CERFACS/CNRS UMR 5318, Toulouse, 31057, France
[2] Collecte Localisation Satellites (CLS), 31520, Ramonville Saint-Agne, France
[3] Centre National d'Etudes Spatiales (CNES), Toulouse, 31401, France

thnguyen@cerfacs.fr

**Abstract**. As the severity and occurrence of flood events tend to intensify with climate change, the need for flood forecasting capability increases. In this regard, the Flood Detection, Alert and rapid Mapping (FloodDAM) project, funded by Space for Climate Observatory initiatives, was set out to develop pre-operational tools dedicated to enabling quick responses in flood-prone areas, and to improve the reactivity of decision support systems. This work focuses on the assimilation of 2D flood extent data (expressed in terms of wet surface ratios) and in-situ water level data to improve the representation of the flood plain dynamics with a Telemac-2D model and an Ensemble Kalman Filter (EnKF). The EnKF control vector was composed friction coefficients and corrective parameter to the input forcing. It is then augmented with the water level state averaged over several floodplain zones. This work was conducted in the context of Observing System Simulation Experiments (OSSE) based on a real flood event occurred in January-February 2021 on the Garonne Marmandaise catchment. This allows to validate the observation operator associated to the wet surface ratio observations as well as the dual state-parameter sequential correction implemented in this work. The merits of assimilating SAR-derived flood plain data complementary to in-situ water level observations are shown in the control parameter and observation spaces with 1D and 2D assessment metrics. It was also shown that the correction of the hydraulic state significantly improved the flood dynamics, especially during the recession. This proof-of-concept study paves the way towards near-real-time flood forecast, making the most of remote sensing-derived flood observations.

**Keywords**. Flood forecasting, hydraulic modeling, data assimilation, ensemble Kalman filter, Sentinel-1, Telemac-2D, observing system simulation experiments.

## 1. Introduction

Floods account for more than 40% of all the globally recorded disasters between 1998 and 2017, according to the United Nations Office for Disaster Risk Reduction (https://www.preventionweb.net/knowledgebase). They are the most common and also the costliest natural disasters worldwide. Yet, quality estimate, assessment, and forecasting of flood hazards remain insufficient in many regions of the world. In order to mitigate flood impacts, hydrodynamic numerical models providing simulations of flood events in analysis and forecast modes are of paramount essential. However, these models remain imperfect due to the uncertainties inherently existing within the models and the inputs, namely friction and boundary conditions (BC), which would be translated into uncertainties in the model outputs.

Flood reanalysis and forecast capability have been greatly improved, owing to the advancements in data assimilation (DA). These methods aim at combining time-series measurements with numerical models to correct the hydraulic states and sequentially reduce the uncertainties in the model parameters [1]. In particular, EnKF relies on stochastically computing the forecast error covariance matrices using a number of member simulations. However, this approach relies strongly on the frequency, density and statistics of errors of the observing network [2]. In reality, limnimetric in-situ gauge stations are only available at a few locations within a catchment [3], usually exclusive in the river bed. This situation can be mitigated by using other data sources such as remote-sensing flood observations which, despite low observation frequency, offer valuable information about the flow dynamics through a 2D representation. In recent years, SAR data has established as a major data source in operational flood management, due to its reliable ability to collect all-day images regardless of weather conditions. Water surfaces and flooded areas typically exhibit low SAR backscatter, as most of the incident radar pulses are specularly reflected away upon arrival at the surfaces. Indeed, flood extent delineation is relatively straightforward on SAR images, with several exceptions, namely built environment and vegetation areas. In our previous works [4, 5], flood extent maps were derived from Sentinel-1 (S1) SAR images by applying a Random Forest classifier. Being used as independent validation observations, they allowed to improve the DA results in both re-analysis and forecast modes.

Taking further advantage of such observations, the assimilation of 2D flood extent maps, represented by wet surface ratios (WSR) provided by the count of wet pixels observed on SAR data, is investigated to comprehensively reduce the inherent uncertainties, and to improve the overall re-analysis and forecast performance. This article presents a novel approach to take into account these WSR observations, jointly with in-situ water-level measurements, within an EnKF framework on top of a 2D hydraulic model. As such, we carry out a dual state-parameter estimation strategy to reduce the uncertainties, namely in friction coefficients, upstream inflow discharge, and hydraulic state in several floodplain areas. This research work is conducted in an OSSE framework, where a reference simulation with pre-determined settings, which is considered as the *truth,* is used to generate synthetical observations. Resulting synthetical in-situ water level and WSR observations are then assimilated into an EnKF with a priori (or *background*) settings which are different from the truth's settings. Such a strategy is quite common in DA studies in order to assess a DA method. It aims at validating whether the proposed method can yield an analyzed control vector that is closer to the truth's settings, as well as providing the resulting model state closer to the observations compared to the a priori information.

## 2. Study Area, Data, Model

### 2.1. Shallow Water Equations (SWEs) in T2D

The hydrodynamic numerical model Telemac-2D (T2D, www.opentelemac.org) is used to simulate and predict water surface elevation and velocity. T2D solves the Shallow Water equations (SWE) derived from depth-averaged Navier-Stokes equations with an explicit first-order time-integration scheme, a finite-element scheme, and an iterative conjugate gradient method [6]. The model results, namely water depth and velocity averaged over the azimuth axis, are provided at each node of the mesh that represents the catchment topo-bathymetry. The non-conservative form of SWEs is written in terms of water level $h$ [m] and velocity $\boldsymbol{u} = (u, v)$ (with horizontal components $u$ and $v$ [m.s$^{-1}$]). Using similar mathematical notations to [4, 7], the SWEs read:

$$\frac{\partial h}{\partial t} + \frac{\partial}{\partial x}(hu) + \frac{\partial}{\partial y}(hv) = 0 \tag{1}$$

$$\frac{\partial u}{\partial t} + \boldsymbol{u} \cdot \nabla u = -g\frac{\partial z}{\partial x} + F_x + \frac{1}{h}\nabla \cdot (h\nu_e \nabla u) \tag{2}$$

$$\frac{\partial v}{\partial t} + \boldsymbol{u} \cdot \nabla v = -g\frac{\partial z}{\partial y} + F_y + \frac{1}{h}\nabla \cdot (h\nu_e \nabla v) \tag{3}$$

where $z$ [m NGF69] is the water surface elevation ($h = z - z_b$ with $z_b$ [m NGF69] being the river bottom elevation) and $\nu_e$ [m$^2$.s$^{-1}$] is the water diffusion coefficient and $g$ [m.s$^{-2}$] is the gravitational acceleration constant. $F_x$ and $F_y$ [m.s$^{-2}$] are the horizontal components of external forces (i.e. friction, wind and atmospheric forces):

$$\begin{cases} F_x = -\dfrac{g}{K_s^2}\dfrac{u|\boldsymbol{u}|}{h^{4/3}} - \dfrac{1}{\rho_w}\dfrac{\partial P_{atm}}{\partial x} + \dfrac{1}{h}\dfrac{\rho_{air}}{\rho_w}C_d U_{w,x}\sqrt{U_{w,x}^2 + U_{w,y}^2} \\ F_y = -\dfrac{g}{K_s^2}\dfrac{v|\boldsymbol{u}|}{h^{4/3}} - \dfrac{1}{\rho_w}\dfrac{\partial P_{atm}}{\partial y} + \dfrac{1}{h}\dfrac{\rho_{air}}{\rho_w}C_d U_{w,y}\sqrt{U_{w,x}^2 + U_{w,y}^2} \end{cases} \quad (4)$$

where $K_s$ [m$^{1/3}$.s$^{-1}$] is the river bed and floodplain friction coefficient according to the Strickler formulation [8], $\rho_w$ and $\rho_{air}$ [kg.m$^{-3}$] are the water and air density, $P_{atm}$ [Pa] is the atmospheric pressure, $C_d$ [-] is the wind drag coefficient that relates the free surface wind to the shear stress, and lastly $U_{w,x}$ and $U_{w,y}$ [m.s$^{-1}$] are the horizontal wind velocity components. In order to solve Eq. (1)-(3), the initial conditions $\{h(x,y,t=0) = h_0(x,y); u(x,y,t=0) = u_0(x,y); v(x,y,t=0) = v_0(x,y)\}$ are required, whereas the upstream BC is described with a time-dependent hydrograph, and a rating curve at the downstream BC.

## 2.2. Study area and description of the uncertainties

The study area, illustrated in Figure 1, is the Garonne Marmandaise catchment which extends over a 50-km reach between Tonneins and La Réole of the Garonne River (southwest France). Three observing stations indicated as black circles are located at Tonneins, Marmande, and La Réole, operated by the VigiCrue network (https://www.vigicrues.gouv.fr/). The white circle indicates a diagnostic location in the floodplain near Marmande; denoted as FPM in the following. A T2D model was developed and calibrated over this catchment [9]. In order to translate the observed water levels into a discharge time-series, the local rating curve at Tonneins is used. This discharge is then applied over the whole upstream interface, covering both river bed and floodplain boundary cells (implemented by EDF R&D [9]), to allow a cold start of the model with any inflow value. However, it prompts a temporary over-flooding of the upstream first meander, until the water returns to the river bed. The downstream BC at La Réole is described in the model with a local rating curve.

Over the simulation domain, the friction coefficient is defined over seven zones, $K_{s_1}$ to $K_{s_6}$ (respective calibrated values: 45, 38, 38, 40, 40 and 40 [m$^{1/3}$.s$^{-1}$]) for the river bed and $K_{s_0}$ (calibrated value 17 [m$^{1/3}$.s$^{-1}$]) for the entire floodplain, as illustrated in Figure 1 with solid colored segments of the rived bed. The initial design of this model by EDF consisted in only three friction segments within the river bed. Based on our previous study [4, 5], a finer zoning definition was favored, which led to the former second and third segments (covering, respectively, the middle and downstream parts of the river bed) were subdivided into two and three smaller segments, respectively $K_{s_{2,3}}$ and $K_{s_{4,5,6}}$. Due to the absence of in-situ data for all of these zones, their a priori values are set based on the calibration process, yet they will allow for a fine adjustment in the DA algorithm. Such a description of the friction coefficients is highly uncertainty-prone due to zoning assumption and calibration procedure. In the following, these coefficients are considered as random variables normally distributed with mean and standard deviation estimated from the calibration process. In addition, the limited number of in-situ observations also yields errors in the formulation of the rating curve that is typically used to translate the observed water-level into discharge, especially for high flow. Therefore, a multiplicative factor $a$ applied on the discharge time-series is considered as a random variable with a normal Probability Density Function (PDF) centered at 1.

Lastly, in order to account for the rainfall, evapotranspiration, and ground infiltration processes that are unavailable in the T2D Garonne model, a correction of hydraulic state in the floodplain has been implemented. Five subdomains (hereafter called *zones*) delineated in the floodplain beyond the dykes, involving a uniform water level correction over each zone, are added to the control vector. These state corrections $\delta H_{[1:5]}$ are considered as zero-mean normal random vector. These zones were determined based on the storage areas, as a result of the dyke system of the catchment [9]. It is worth-noting that the first storage area of the model, at the first meander near Tonneins, is excluded in this study because of the aforementioned artificial temporary over-flooding effect. In addition, several storage areas near the downstream area are not considered, because these areas are not fully observed by S1, and spurious dynamics may be caused by the errors in topography [9]. Over each of the five zones, the wet surface ratio (WSR) between the area of observed wet surfaces and the total area of the zone is measured.

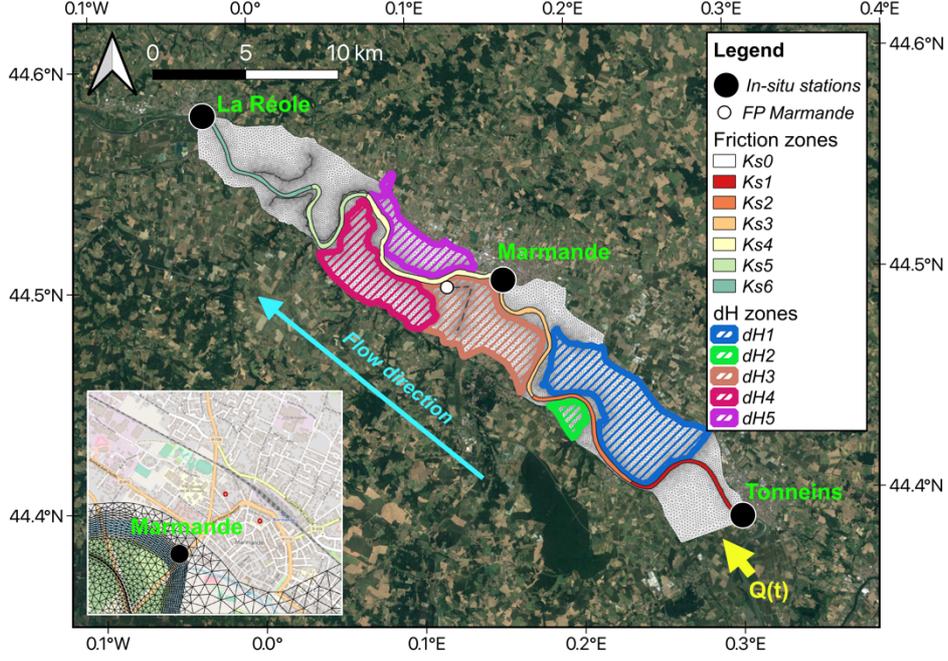

**Figure 1.** T2D Garonne Marmandaise model and control vector. The inset figure shows the impacted urban area near Marmande. (Note: from Nguyen et al. (2022), Dual state-parameter assimilation of SAR-derived wet surface ratio for improving fluvial flood reanalysis, *Water Resources Research*, 58, e2022WR033155. CC BY-NC.)

The proof-of-concept of the DA was carried out for a substantial flood event that occurred on the Garonne catchment in late January and February 2021. This event exceeded the highest threshold level for flood risks, set out by the French national flood forecasting center (SCHAPI) in collaboration with the departmental prefect. The flood peak was reached on 2021-02-04. In this work, we examine an extended length of this event between 2021-01-16 and 2021-02-10, during which the catchment was observed nine times by S1 images.

## 3. Data Assimilation

### 3.1. Description of the control vector

The DA algorithm consists in a cycled stochastic EnKF, where the control vector **x** is composed of the seven friction coefficients $K_{s_{[0:6]}}$, one multiplicative parameter $a$ to modify the time-varying upstream BC, and five state corrective variables $\delta H_{[1:5]}$ over the floodplain zones. These $n = 13$ parameters are assumed to be constant within a DA cycle, and allowed by the DA to evolve from one cycle to another. Each cycle $k$ covers a time window, denoted by $W_k = [t_{start}, t_{end}]$ of 18-hour length over which $N_{obs}$ observations are assimilated. The cycling consists in sliding the windows by $T_{shift} = 6$ hours so that two consecutive cycles $W_k$ and $W_{k+1}$ overlap for 12 hours. This EnKF algorithm relies on the propagation of $N_e$ members with perturbed values of **x**, i.e. the forecast values denoted by $\mathbf{x}_k^{f,i}$ with $i \in [1, N_e]$ represents the ensemble member index.

### 3.2. Forecast step

The forecast step stands in the propagation in time of the control and model state vectors, whereas the analysis step (section 3.3) stands in the update of these vectors. However, the EnKF is here applied to model parameters (in the control vector **x**) that do not evolve during each $W_k$. Artificial dispersion is further introduced when adding perturbations $\boldsymbol{\theta}_k^i$ to the mean of the analysis from the previous cycle $\overline{\mathbf{x}_{k-1}^a}$ to avoid ensemble collapse. The forecast step thus reads:

$$\mathbf{x}_k^{f,i} = \begin{cases} \mathbf{x}_0 + \boldsymbol{\theta}_1^i & \text{if } k = 1 \\ \overline{\mathbf{x}_{k-1}^a} + \boldsymbol{\theta}_k^i & \text{if } k > 1 \end{cases} \quad (5)$$

with $\overline{\mathbf{x}^a_{k-1}} = \left(\sum_{i=1}^{N_e} \mathbf{x}^{a,i}_{k-1}\right)/N_e \in \mathbb{R}^n$ and $\boldsymbol{\theta}^i_k \sim \mathcal{N}(\mathbf{0}, \sigma^2_{i,k})$ where

$$\sigma_{i,k} = \begin{cases} \sigma_{\mathbf{x}} & \text{if } k = 1 \\ \lambda \sqrt{\frac{1}{N_e}\sum_{i=1}^{N_e}\left(\mathbf{x}^{a,i}_{k-1} - \overline{\mathbf{x}^a_{k-1}}\right)^2} + (1-\lambda)\sigma_{\mathbf{x}} & \text{if } k > 1 \end{cases} \quad (6)$$

The controlled variables issued from $\overline{\mathbf{x}^a_{k-1}}$ is further dispersed with perturbations $\boldsymbol{\theta}^i_k$ drawn from the zero-mean normal distribution with the standard deviation obtained from the linear combination (weighted by $\lambda = 0.3$) of the $\sigma_{\mathbf{x}}$ and the standard error of the analysis at previous cycle. This technique is an alternative for anomaly inflation in order to avoid the ensemble from collapsing, while preserving part of the information from the background statistical characteristics. The corrections to the hydraulic state variable $\delta H_{[1:5]}$ are drawn within zero-mean normal distributions, thus fully described by perturbations $\boldsymbol{\theta}^i_k$ (i.e. without the mean of analysis from previous cycle).

The background hydraulic state associated with each ensemble member, denoted by $\mathbf{s}^{f,i}_k$, results from the integration of the hydrodynamic model $\mathcal{M}_k: \mathbb{R}^n \to \mathbb{R}^m$ from the control space to the model state space over $W_k$: $\mathbf{s}^{f,i}_k = \mathcal{M}_k(\mathbf{s}^{a,i}_{k-1}, \mathbf{x}^{f,i}_k)$. For the very first cycle, the initial condition for $\mathcal{M}_k$ at $t_{start}$ is provided by a restart file defined by user. Then, for the following cycles $k$, it requires in a full restart from $\mathbf{s}^{a,i}_{k-1}$, saved from the analysis run of the previous cycle $\mathbf{s}^{a,i}_{k-1} = \mathcal{M}_{k-1}(\mathbf{s}^{a,i}_{k-2}, \mathbf{x}^{a,i}_{k-1})$ at the beginning of each $W_k$. In order to ensure the consistencies between the hydraulic state and the analyzed parameters at $t_{start}$, a short (3-hour preceding $t_{start}$) spin-up integration can be carried out.

The control vector equivalent in the observation space for each member, noted $\mathbf{y}^{f,i}_k$, stems from:

$$\mathbf{y}^{f,i}_k = \mathcal{H}_k(\mathbf{s}^{f,i}_k) - \mathbf{y}_{bias} \quad (7)$$

where $\mathcal{H}_k: \mathbb{R}^m \to \mathbb{R}^{N_{obs}}$ is the observation operator from the model state space to the observation space that selects, extracts and eventually interpolates model outputs (if necessary) at times and locations of the observation vector $\mathbf{y}^o_k$ over $W_k$. In this work, the observation operator also includes a correction to account for a systematic model-observation bias, denoted by $\mathbf{y}_{bias}$. Such a bias was estimated based on the 24-hour quasi-stationary non-overflowing period of 2021-01-15, prior to the real flood event in January-February 2021 [4].

### 3.3. Analysis step

When applying a stochastic EnKF [10], the observation vector $\mathbf{y}^{o,i}$ is perturbed, thus an ensemble of observations $\mathbf{y}^{o,i}_k$ ($i \in [1, N_e]$) is generated:

$$\mathbf{y}^{o,i}_k = \mathbf{y}^o_k + \boldsymbol{\epsilon}_k \text{ with } \boldsymbol{\epsilon}_k \sim \mathcal{N}(0, \mathbf{R}_k) \quad (8)$$

where $\mathbf{R}_k = \sigma_{obs}^2 \mathbf{I}_{N_{obs}}$ is the observation error covariance matrix, assumed to be diagonal, of standard deviation $\sigma_{obs}$ (and $\mathbf{I}_{N_{obs}}$ is the $N_{obs} \times N_{obs}$ identity matrix). For the in-situ water level observation, the standard deviation is proportional to the observations $\sigma_{obs,k} = \tau \mathbf{y}^o_k$, whereas the $\sigma_{obs}$ takes fixed values for the WSR observations. The innovation vector (over $W_k$) computes the difference between the perturbed observation vector $\mathbf{y}^{o,i}_k$ (from Eq. (7)) and the model equivalent $\mathbf{y}^{f,i}_k$ (from Eq. (8)). Such vector is weighted by the Kalman gain matrix $\mathbf{K}_k$ and then added as a correction to the background control vector $\mathbf{x}^{f,i}_k$, so that the analysis control vector $\mathbf{x}^{a,i}_k$ is as follows,

$$\mathbf{x}^{a,i}_k = \mathbf{x}^{f,i}_k + \mathbf{K}_k(\mathbf{y}^{o,i}_k - \mathbf{y}^{f,i}_k) \quad (9)$$

The Kalman gain reads: $\mathbf{K}_k = \mathbf{P}^{x,y}_k [\mathbf{P}^{y,y}_k + \mathbf{R}_k]^{-1}$ with $\mathbf{P}^{y,y}_k$ being the covariance matrix of the errors in the background state equivalent in the observation space $\mathbf{y}^f_k$ and $\mathbf{P}^{x,y}_k$ the covariance matrix between the errors in the control vector and the error in $\mathbf{y}^f_k$. They are stochastically estimated within the ensemble:

$$\mathbf{P}^{x,y}_k = \frac{1}{N_e} \mathbf{X}^T_k \mathbf{Y}_k \in \mathbb{R}^{n \times N_{obs}} \quad \text{and} \quad \mathbf{P}^{y,y}_k = \frac{1}{N_e} \mathbf{Y}^T_k \mathbf{Y}_k \in \mathbb{R}^{N_{obs} \times N_{obs}} \quad (10)$$

with $\mathbf{X}_k = \left[\mathbf{x}_k^{f,1} - \overline{\mathbf{x}_k^f}, \cdots, \mathbf{x}_k^{f,N_e} - \overline{\mathbf{x}_k^f}\right] \in \mathbb{R}^{n \times N_e}$ and $\mathbf{Y}_k = \left[\mathbf{y}_k^{f,1} - \overline{\mathbf{y}_k^f}, \cdots, \mathbf{y}_k^{f,N_e} - \overline{\mathbf{y}_k^f}\right] \in \mathbb{R}^{N_{obs} \times N_e}$ where $\overline{\mathbf{x}_k^f} = \frac{1}{N_e}\sum_{i=1}^{N_e} \mathbf{x}_k^{f,i} \in \mathbb{R}^n$ and $\overline{\mathbf{y}_k^f} = \frac{1}{N_e}\sum_{i=1}^{N_e} \mathbf{y}_k^{f,i} \in \mathbb{R}^{N_{obs}}$.

The analyzed hydrodynamic state $\mathbf{s}_k^{a,i}$, associated with each analyzed control vector $\mathbf{x}_k^{a,i}$ is denoted by $\mathbf{s}_k^{a,i}$. It results from the integration of the hydrodynamic model $\mathcal{M}_k$ with the updated controlled parameters over $W_k$, starting from the same initial condition (for the first cycle), then from each background simulation within the ensemble: $\mathbf{s}_k^{a,i} = \mathcal{M}_k(\mathbf{s}_{k-1}^{a,i}, \mathbf{x}_k^{a,i})$.

## 4. Observing System Simulation Experiment (OSSE)

### 4.1. Truth and observation generation

The framework of an OSSE consists in a deterministic simulation with a chosen set of parameters, which are time-varying in the present case. This reference simulation is considered the *truth*. In the present work, the so-called true friction parameters and inflow correction are set from two previous DA experiments on the real 2021 flood event that assimilate in-situ observations and WSR observations. The state correction true values were set up with negative cosine curves at the three first groups of S1 observations, and then with a constant negative water of -18 cm at the recession period. The *synthetic observations* are generated based on the reference simulation. They are then assimilated in a DA experiment, with initial calibrated/default values that are different from the truth's settings. The upstream time-varying BC at Tonneins is also provided by the real flood event in January-February 2021. However, only synthetical observations are assimilated. It is worth-noting that the corrections of the hydraulic state in the floodplain ($\delta H_{[1:5]}$) are only performed at the assimilation window in which WSR observations are present. A full study including real flood events can be found in [11].

### 4.2. Experimental setup

One free run and three DA experiments were carried out (Table 1) with different types of observations that are assimilated and the different components of the control vector. Two types of observations are concerned: *(i)* in-situ water-level observations at three observing stations Tonneins, Marmande, and La Réole, *(ii)* WSR measurements on the five floodplain zones (associated with $\delta H_{[1:5]}$). For the DA experiments, the value of $\tau$ for in situ data is fixed to 15%, meaning that $\sigma_{obs}$ amounts to 15% of the observation value, whereas the value of $\sigma_{obs}$ associated with WSR data varies from 0.2 down to 0.1 according to when the observation time is located within the 18-hour assimilation window.

**Table 1.** Summary of the experiment settings.

| Exp. name | FR or DA? | Assimilated observations | $N_e$ | Control vector |
|---|---|---|---|---|
| FR | FR | - | 1 | - |
| IDA (in-situ DA) | DA | In-situ water level | 75 | $K_{S_0}, K_{S_{[1:6]}}, a$ |
| IWDA | DA | In-situ water level & WSR | 75 | $K_{S_0}, K_{S_{[1:6]}}, a$ |
| IHDA | DA | In-situ water level & WSR | 75 | $K_{S_0}, K_{S_{[1:6]}}, a, \delta H_{[1:5]}$ |

## 5. Results and Discussions

Quantitative performance is assessed in the control and observation spaces. They involve the comparisons between *(i)* the analyzed parameters and the truth's parameters, *(ii)* the analyzed water-level time-series and synthetical in-situ observations, *(iii)* the analyzed WSR and synthetical WSR observations in the floodplain, and lastly the evaluation of *(iv)* the overall Critical Success Index (CSI) computed over the analyzed and the truth's flood extent maps.

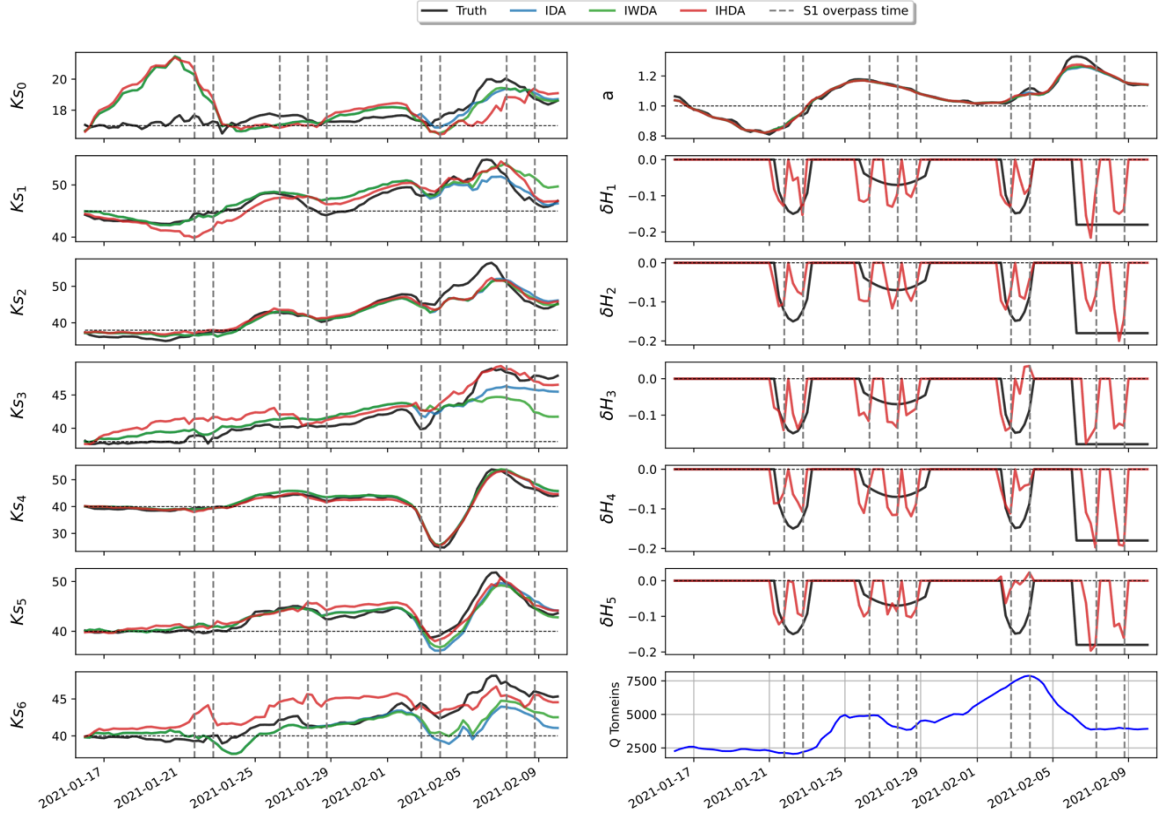

**Figure 2.** Evolution of controlled parameters for friction, multiplicative correction applied on the inflow, water level correction in the floodplain zones. Last panel depicts the forcing inflow at Tonneins.

## 5.1. Results in the control space

Figure 2 shows the analyzed parameters from the DA experiments, with IDA represented by the blue curves, IWDA by green, and IHDA by red ones. In each panel, the *truth* parameters are depicted in black, whereas the calibrated and default values $\mathbf{x}_0$ are shown by the horizontal dashed lines. The overpass times of S1 in this event are depicted by the vertical dashed lines. It appears that for most of riverbed friction zones (except $K_{S_6}$), the three DA analyses are able to retrieve the true value of the friction coefficients. Among them, IHDA (red curves) with the extended control vector provides the closest estimation of the *truth*. In the floodplain, the correction of the friction $K_{S_0}$ exhibits some difficulty for all DA experiments and it takes several days to converge to the true parameter, despite the assimilation of WSR observations. It should be noted that in spite of slow convergence or equifinality issues, all analyzed friction coefficients remain within physically feasible ranges. All DA experiments provide excellent results for the inflow multiplicative factor $a$, except near the flood peak. Since the analysis for the state correction in the floodplain $\delta H_{[1:5]}$ is only activated when WSR observations are present over the assimilation window, starting from 18 hours before each observation, hence the state correction is null-valued for most of the time, including in between S1 overpass times (minimum 24 hours between two observations). Yet, the correction reaches values that are close to the true values.

## 5.2. Results in the observation space: Water levels at observing stations

The dynamic of the river bed and the floodplain is assessed with respect to in situ data at Tonneins, Marmande, La Réole, and FPM location (not assimilated) respectively, by the Root-Mean-Square Errors (RMSE) computed between the reference water level time-series and the FR/DA experiments. For each column, the lowest RMSEs are boldfaced. Table 2 shows that all DA experiments succeed in significantly reducing the water level errors, compared to that of FR. The DA experiments RMSE at the three observing stations, where water lever synthetical data are assimilated remains under 5.3 cm. This level of precision can be expected in OSSE, as it validates the performance of the EnKF algorithm even though the ensemble size is limited (with $N_e = 75$ members). The RMSE reduction with respect to FR amounts to 80%, 89%, and 91%, respectively, at Tonneins, Marmande, and La Réole, while is reaches 34.4% (for IDA and IWDA) or 61.1% (for IHDA) at FPM where no data is assimilated. The merits of correcting the water level state are shown as IHDA outperforms the other two DA experiments, especially at FPM.

**Table 2.** Water level RMSE w.r.t. in-situ water levels at VigiCrue observing stations and FPM.

| | RMSE [m] | | | |
|---|---|---|---|---|
| *Exp. name* | *Tonneins* | *Marmande* | *La Réole* | *FPM (only validation)* |
| FR | 0.259 | 0.393 | 0.572 | 0.483 |
| IDA | **0.051** | **0.043** | 0.051 | 0.317 |
| IWDA | 0.053 | 0.044 | 0.050 | 0.317 |
| IHDA | **0.051** | **0.043** | **0.048** | **0.188** |

### 5.3. Results in the observation space: WSR in the floodplain

The WSR in the five floodplain zones for FR and DA are compared to the reference values, as shown in Figure 3 with the same color code as in Figure 2. The difference between the WSR simulated by FR and DA experiments and the reference *true* are shown. Up to the water level rise (around 2021-02-01), the impact of DA is not significant as the floodplain is not active. At the flood peak, the FR (orange curves) over-predicts the flooding in zone 1, while under-predicting highly the flooding in zones 2, 3, and 4. Same results are obtained with IDA and IWDA in zone 1 at the flood peak, yet both of them allow for an improvement in zones 2-4 around the flood peak. The correction of the water level state in IHDA allows for a significant improvement, especially during the flood recession when the T2D model alone fails to evacuate the water. It appears that the augmented control vector together with the assimilation of WSR data is an efficient strategy that allows to account for model error and leads to the best results.

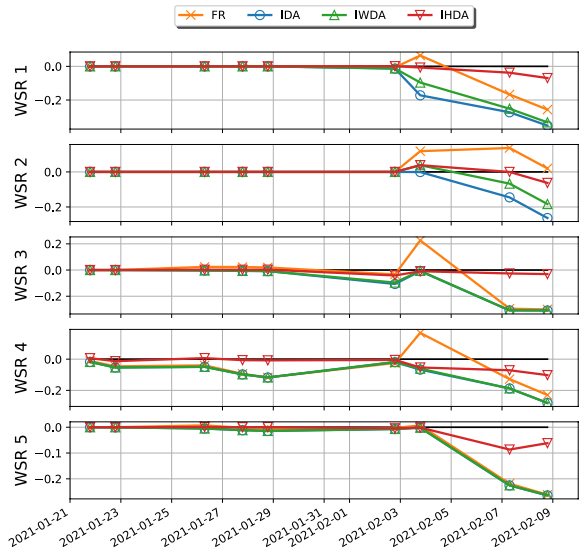

**Figure 3 (left).** Errors between truth's WSR values and simulated WSR values in the 5 floodplain zones.

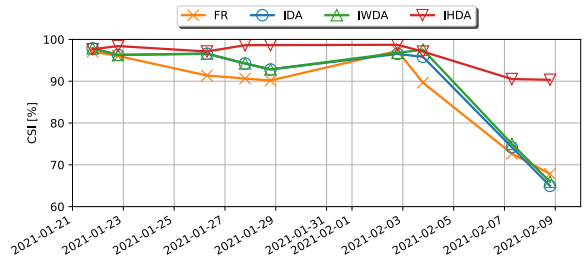

**Figure 4.** CSI scores between the realized experiments.

### 5.4. 2D validation with Critical Success Index (CSI)

The CSI is computed to validate the agreement between the flood extent maps predicted by the FR (and DA) experiments with respect to those based on the *truth* simulation (i.e., the reference flood extent maps). This metric relies on the pixel counts combining the four possible outcomes: the number of pixels correctly predicted as flooded and non-flooded, the number of non-flooded pixels incorrectly predicted as flooded (or *over-prediction*), and the number of the flooded pixels being missed by the prediction (or *under-prediction*). Figure 4 summarizes the CSI yielded by all four performed experiments at every S1 overpass times. It should be noted that all the resulting flood extent maps from DA experiments are in relative agreements with the reference flood maps. As such, all three DA experiments present a significant improvement near the flood peak (2021-02-03 19:00) compared to FR. In particular, IHDA results in CSI above 90% at every S1 overpass times (and above 96% before recession period). The performance is shown to be better at various moments during the whole flood event, especially during the water recession period where the other two DA experiments IDA and IWDA yield a low CSI (65-75%).

# 6. Conclusions and Perspectives

This work studies the assimilation of flood extent maps, represented by WSR, to improve the hydrodynamic simulation results based on numerical T2D models with EnKF. The study was carried out over the Garonne Marmandaise catchment, focusing on the a synthetical flood event. Four experiments were realized, namely one in free run mode and three in DA mode in an OSSE framework. All of the DA experiments were implemented by a cycled EnKF with an 18-hour assimilation window sliding with 12-hour overlapping. This study confirms the assertion that a densification of the observing network allows to mitigate the equifinality issues and improve the flood forecast capability. These first two DA experiments (IDA and IWDA) focus on sequentially correcting friction coefficients and inflow discharge (through a multiplicative factor). However, IDA assimilates only in-situ water-level observations whereas IWDA involves jointly in-situ water level and WSR observations derived from SAR flood extent maps. They are shown to be capable of retrieving the true controlled parameters, thus providing relevant results. Nevertheless, this article highlights particularly the merits of IHDA experiment, which also assimilates both types of observations. In addition, it deals with friction coefficients and inflow discharge as well as the hydraulic state variable in five determined floodplain zones, representing rainfall, evapotranspiration and/or ground infiltration processes, based on a dual state-parameter estimation within the EnKF. This approach is able to result in simulated water levels and WSRs that are very close to the synthetic observations, as well as yielding better estimates of true friction and discharge parameters, compared to IDA and IWDA. In terms of 1D assessment, IHDA presents similar RMSE to IDA and IWDA (if not slightly better) on the three observing stations. The largest error on both three DA experiments is only 5.3 cm, which confirms a very high precision for the performed cycled DA approach. Nevertheless, IHDA results in a water level RMSE on the floodplain which is reduced 40.7%, compared to the RMSEs provided by IDA and IWDA. The 2D assessment also shows the advantage of the IHDA over the other two DA experiments, in particular before the flood peak and during the recession period, highlighted by a CSI always above 90% (and above 96% before the recession).

Moreover, this article has validated the developed observation operator dedicated to assimilating WSR observations, as well as the appropriate augmented control vector. These observations were able to be accommodated into the EnKF paradigm without conflicts with in-situ observations. Yet, it should be stressed that the selection of floodplain zones to be corrected with $\delta H$ is quite a challenging task, which involved crucial understanding of the T2D model. In addition, within this research work, we have also carried out several different approaches concerning the localization within EnKF, meaning the isolation of the influences of WSR observations onto friction coefficients and inflow discharge, as well as the impact of in-situ observation on the $\delta H$ variables. However, since the resulting effects are quite insignificant for a sufficiently large ensemble, this article only presents the experiments and results with 75 members and without any localization.

A particular perspective of this work concerns the non-gaussianity of these bounded WSR observations. This can lead to a suboptimality of the EnKF algorithm, since it requires the condition that observational errors follow normal distributions. Several existing studies have been investigated on a variable change, widely known as Gaussian anamorphosis, in order to realize the analysis step in a Gaussian space. Furthermore, we aim to tackle the subject of the assimilation of SAR-derived flood extent maps in near real time, expressed similarly in WSR measurements. In addition, the proposed dual state-parameter estimation approach can also be applied to correct the highly-uncertain topography over the downstream areas. Lastly, we are also interested in expanding the observing network from space with other types of data, namely a higher resolution SAR dataset (TerraSAR-X) and optical imagery dataset (Sentinel-2).

**Acknowledgement**
This work was supported by CNES, CERFACS and SCO-France. The authors gratefully thank the EDF R&D for providing the Telemac-2D model for the Garonne River, and the SCHAPI, the SPC Garonne-Tarn-Lot and the SPC Gironde-Adour-Dordogne for providing the in-situ data.